\documentclass{elsart}

 \usepackage{graphicx}
  \usepackage{epsfig}

\usepackage{amssymb}

\begin{document}

\begin{frontmatter}

\title{Infrared extinction by homogeneous particle aggregates of 
SiC, FeO and SiO$_2$: comparison of different theoretical approaches}

\author[1,2]{Anja C. Andersen}
\author[3]{Harald Mutschke}
\author[4,3]{Thomas Posch}
\author[5]{Michiel Min}
\author[3]{Akemi Tamanai}
\address[1]{Dark Cosmology Center, Juliane Maries Vej 30, DK-2100 Copenhagen, Denmark}
\address[2]{NORDITA, Blegdamsvej 17, DK-2100 Copenhagen, Denmark} 
\thanks{phone: (+45) 3532 5229, fax: (+45) 3538 9157 , e-mail: anja@nordita.dk }
\address[3]{Astrophysikalisches Institut, Schillerg\"asschen 2-3, D-07745 Jena, Germany}
\address[4]{Institut f\"ur Astronomie, T\"urkenschanzstra{\ss}e 17, A-1180 Wien, Austria}
\address[5]{Astronomical Institute Anton Pannekoek, University of Amsterdam, Kruislaan 403, NL-1098 SJ Amsterdam, The Netherlands}

\begin{abstract}
Particle shape and aggregation have a strong influence on the spectral profiles 
of infrared phonon bands of solid dust grains. Calculating these effects 
is difficult due to the often extreme refractive index values in these bands. 
In this paper, we use the Discrete Dipole Approximation (DDA) and the 
T-matrix method to compute the absorption band profiles for simple clusters of touching 
spherical grains. We invest reasonable amounts of computation time in order to 
reach high dipole grid resolutions and take high multi-polar orders into account, 
respectively. 
The infrared phonon bands of three different refractory materials of astrophysical 
relevance are considered - Silicon Carbide (SiC), Wustite (FeO) and Silicon 
Dioxide (SiO$_2$). We demonstrate that even though these materials display a 
range of material properties and therefore different strengths of the 
surface resonances, a complete convergence is obtained with none of the approaches. 
For the DDA, we find a strong dependence of the calculated band profiles on the 
exact dipole distribution within the aggregates, especially in the vicinity of the 
contact points between their spherical constituents. 
By applying a recently developed method to separate the material optical constants 
from the geometrical parameters in the DDA approach, we are able to demonstrate 
that the most critical material properties are those where the real part of the 
refractive index is much smaller than unity. 

\end{abstract}
\end{frontmatter}

Grain growth by aggregation is an important process in dense cosmic 
environments as well as in the Earth's atmosphere. Besides influencing 
dynamic properties it also changes the absorption and scattering properties 
of the solid dust particles for electromagnetic radiation (e.g.\ 
\cite{min+etal05a,stog95}). This is especially true in spectral regions where resonant 
absorption occurs, such as the phonon bands in the infrared (IR). It is 
quite well known that shape and aggregation effects actually determine 
the band profiles of such absorption and emission bands, which hinders 
e.g.\ the identification of particulate materials by their IR bands. 
Despite many detailed investigations the full understanding of the 
influence of grain aggregation is still lacking, due to the complicated
nature of the problem (e.g.\ \cite{andersen+etal02,smith+etal95}). 

We have investigated different simplified cluster shapes, as a means to
compare how the numerical codes which are often used in astrophysical 
calculations manage to converge. The clusters consist of 5, 7 and 8 particles and are 
of significantly different shapes. For the material properties we have considered
three astrophysically relevant materials differing in their optical properties; Silicon
Carbide (SiC), Wustite (FeO) and Silicon Dioxide (SiO$_2$).

The paper is  structured such that Sect.\,\ref{structure_sect} presents a 
description of the cluster shapes, Sect.\,\ref{materials_sect} provides a 
description of the different materials considered and their optical properties. 
In Sect.\,\ref{codes_sect} the numerical methods used for determining the 
extinction of the clusters are described. Sections \ref{sect_cos} and 
\ref{sect_DDA} describe our results for the cluster-of-spheres method and the
DDA calculations, respectively, and describe the convergence problems 
encountered by the different methods. 
Sect.\,\ref{conclusion_sect} offers our conclusions.

\section{Cluster Structures} \label{structure_sect}

\begin{figure}
\centering\includegraphics[width=9cm,angle=0]{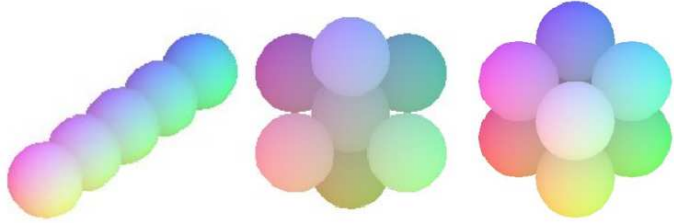}
\label{clusters}
\caption{The clusters we study; a linear cluster of five particles {\it lin5},
a seven particle cluster {\it frac7} and a eight particle cluster {\it sc8}. }
  \hfill
\end{figure}

We consider clusters of identical touching spherical 
particles arranged in three different geometries: fractal, cubic, and linear. 
For a high precision of the calculations, we restrict the number of 
particles per cluster to less than 10.  

We have selected  
three main geometries, namely a ``snowflake 1'st--order prefractal'' cluster 
(fractal dimension $D=\ln7/\ln3=1.77$, {\normalsize \cite{vicsek83}}), 
where one sphere is surrounded by six others along the positive and negative 
cartesian axes, a cluster of eight spheres arranged as a cube and 
a linear chain of five spheres. All of the clusters (Fig.\,\ref{clusters}) 
consist of homogeneous spheres with radii $R=10$~nm and embedded in vacuum 
(or air). 

\begin{figure}
\centering\includegraphics[width=7cm,angle=0]{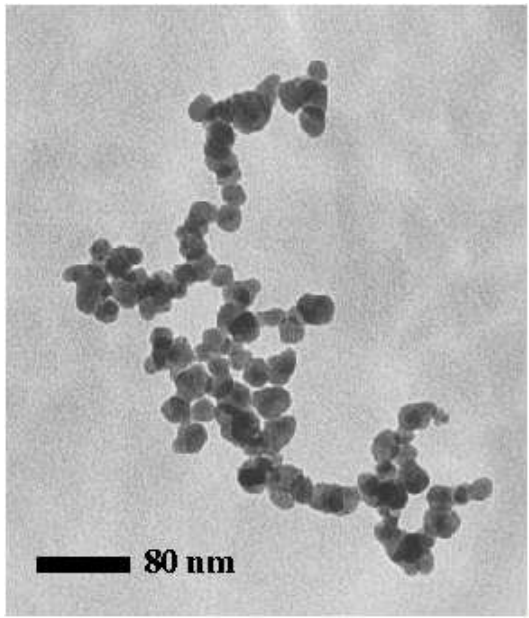}
\label{clement}
\caption{Transmission Electron Micrograph of an $\beta$-SiC cluster produced
in the laboratory by laser-induced pyrolysis \cite{clement03}. }
  \hfill
\end{figure}

An SiC cluster consisting of nano-particles produced in the 
laboratory by laser-induced pyrolysis \cite{clement03}, which has served as 
inspiration for this paper, is shown in Fig.\,\ref{clement}. 

\section{Infrared properties of our materials}\label{materials_sect}

 \begin{figure}
\resizebox{\hsize}{!}{\includegraphics{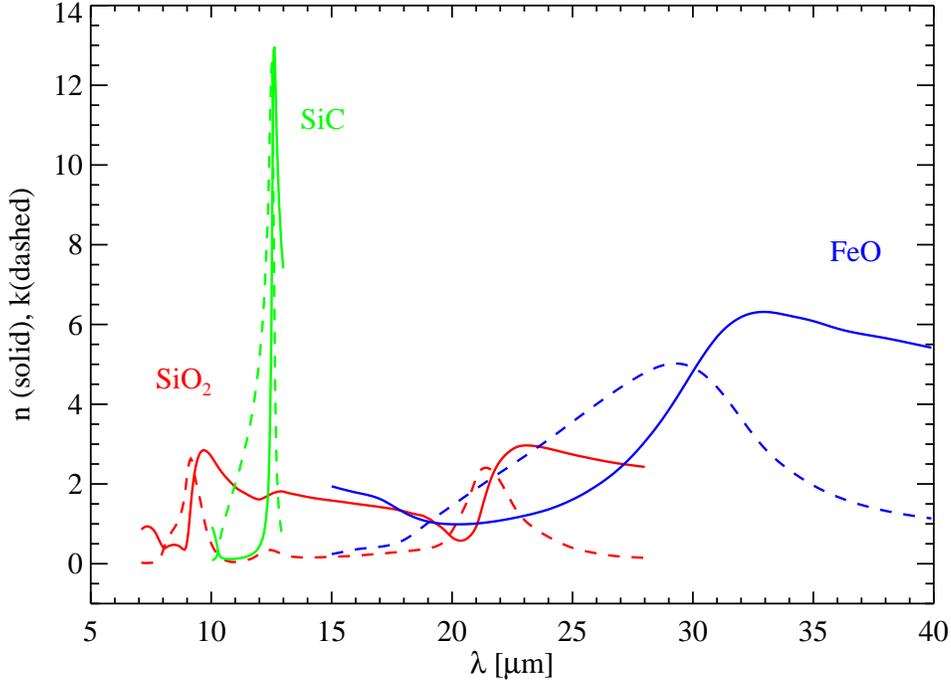}}
\label{nk}
 \caption{The optical constants n and k for the three different materials
 considered for the clusters; SiC, FeO and SiO$_2$. It can be seen that SiC
 possesses a significantly stronger resonances than the two other materials. }
 \end{figure}

In order to study the convergence of the different codes, we chose three
different materials of which our clusters are composed: crystalline
$\beta$-SiC, crystalline FeO and amorphous SiO$_2$. These solids differ
significantly from each other in their infrared properties. First, while
$\beta$-SiC and FeO have only one mid-infrared resonance, amorphous SiO$_2$
has three of them. Second -- and more importantly -- the minima and peak
values of these materials' optical constants are very different from each other,
see Fig.\,\ref{nk}.
Roughly speaking, the maximum $n$ and $k$ values of FeO, by reaching 5--6
close to 30\,$\mu$m, amount to only 50\% of those of $\beta$-SiC in its resonance 
at 12.6\,$\mu$m, and the
same holds true for SiO$_2$ as compared to FeO. Due to its sharp resonance,
$\beta$-SiC not only reaches a very high maximum, but also a very deep
minimum
of $n$, which makes the calculation of its absorption and scattering
behaviour
especially difficult (see also Sect.\,\ref{sect_cos} and \ref{sect_DDA}). 
More details on the maxima
of $n$ and $k$ as well as the minima of $n$ for SiC, FeO and SiO$_2$
can be found in Tab.\ \ref{t:nk-max}.

\begin{table}[|htbp|]
\begin{center}
         \caption{Maxima of the optical constants $n$ and $k$ of SiC,
         FeO and SiO$_2$. Amorphous SiO$_2$ has three resonances in
         the mid-IR range which have been labeled r1 \dots r3.
         \label{t:nk-max}} 
\begin{tabular}{|lcccccc|}
\hline
Resonance & n$_{max}$ & $\lambda$(n$_{max}$) & n$_{min}$ & $\lambda$(n$_{min}$) & k$_{max}$ & $\lambda$(k$_{max}$) \\
 &  & [$\mu$m] &  & [$\mu$m]  & & [$\mu$m]  \\ \hline \hline
$\beta$-SiC & 13.0 & 12.6  & 0.12 & 10.8 & 12.6  & 12.5  \\
FeO         & 6.3  & 33.0  & 0.99  & 20.3 & 5.0   & 29.2  \\
SiO$_2$, r1 & 2.8  & 9.8  & 0.36 &  8.9 & 2.6   & 9.2   \\
SiO$_2$, r2 & 1.8  & 12.8 & 1.62 & 11.9 & 0.3   & 12.4  \\
SiO$_2$, r3 & 2.9  & 22.7  & 0.58 & 20.4 & 2.4   & 21.3  \\ \hline
\end{tabular} 
\end{center}
\end{table}

The optical constants used for the amorphous SiO$_2$ we got from
\cite{palik85} and for FeO from \cite{henning+etal95}.
The data for $\beta$-SiC were calculated from a Lorentz oscillator model
according to \cite{mutschke99}, assuming a damping constant, $\gamma$, of
10\,cm$^{-1}$. It should be noted that high-quality (terrestrial)
SiC is characterized by damping constants of only 1--3\,cm$^{-1}$,
implying an even sharper resonance than for $\gamma$ = 10\,cm$^{-1}$.
However, circumstellar SiC is almost certainly characterized by an
imperfect lattice structure, corresponding to a damping constant
significantly larger than 3\,cm$^{-1}$ (see \cite{mutschke99} for
more details).

\begin{figure}
\resizebox{\hsize}{!}{\includegraphics{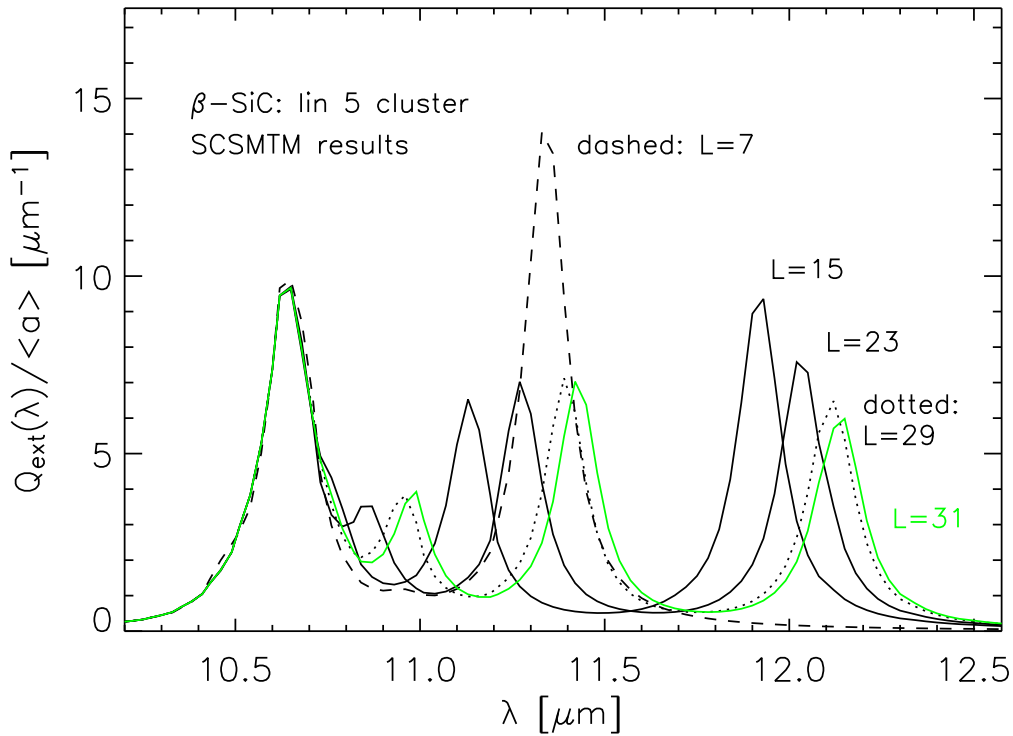}}
\caption{The calculated extinction of a SiC linear five particle cluster {\it lin5} using the T-matrix SCSMTM code by \cite{mackowski96}. The polar order, L, of each calculation is indicated by an annotation.
}
\label{lin5_SiC_SCSMTM}
\end{figure}

SiC, FeO and SiO$_2$ are not only among the potential condensates
in circumstellar shells, but due to the above indicated values of their
infrared optical constants also well-suited for testing the convergence of
the different codes delivering the cluster absorption and scattering
effiencies.
As we are going to show, covergence in the absorption efficiency calculations
cannot be reached by means of the codes we used for materials with optical
constants reaching as deep minima and as high maxima of $n$ as $\beta$-SiC
does; but also for a material like amorphous SiO$_2$ the convergence of
our calculations is very slow.

\section{Computational approaches}{\label{codes_sect}}

\begin{figure}
\resizebox{\hsize}{!}{\includegraphics{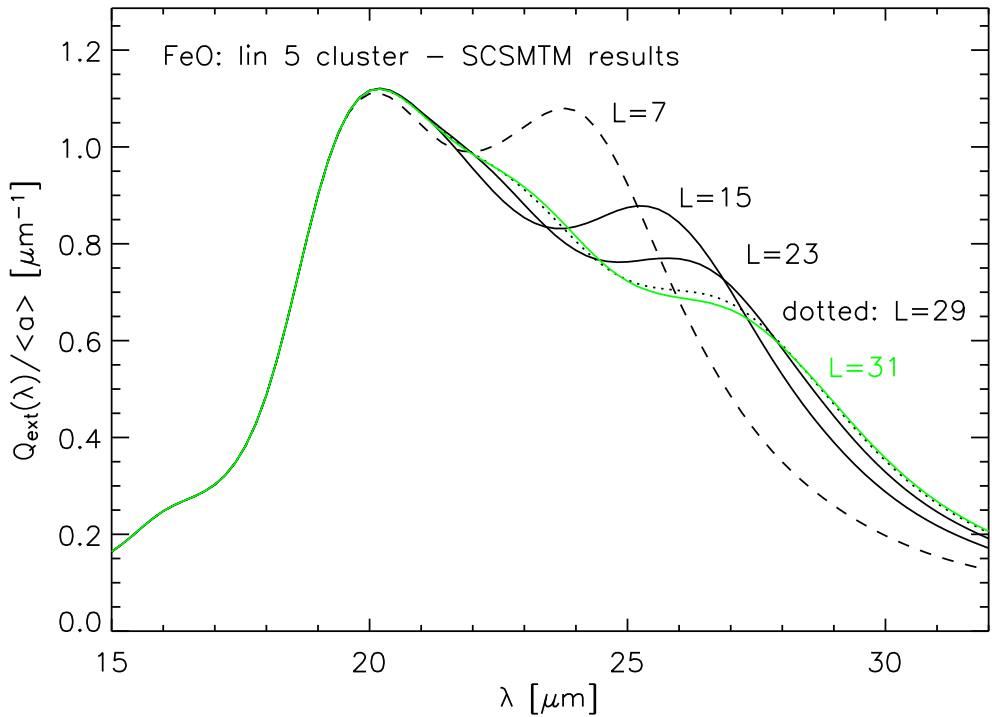}}
\caption{The calculated extinction of a FeO linear five particles cluster {\it lin5} using the SCSMTM code \cite{mackowski96}. The polar orders shown span from L is 7 to 31.}
\label{FeO_lin5_SCSMTM}
\end{figure}

\begin{figure}
\resizebox{\hsize}{!}{\includegraphics{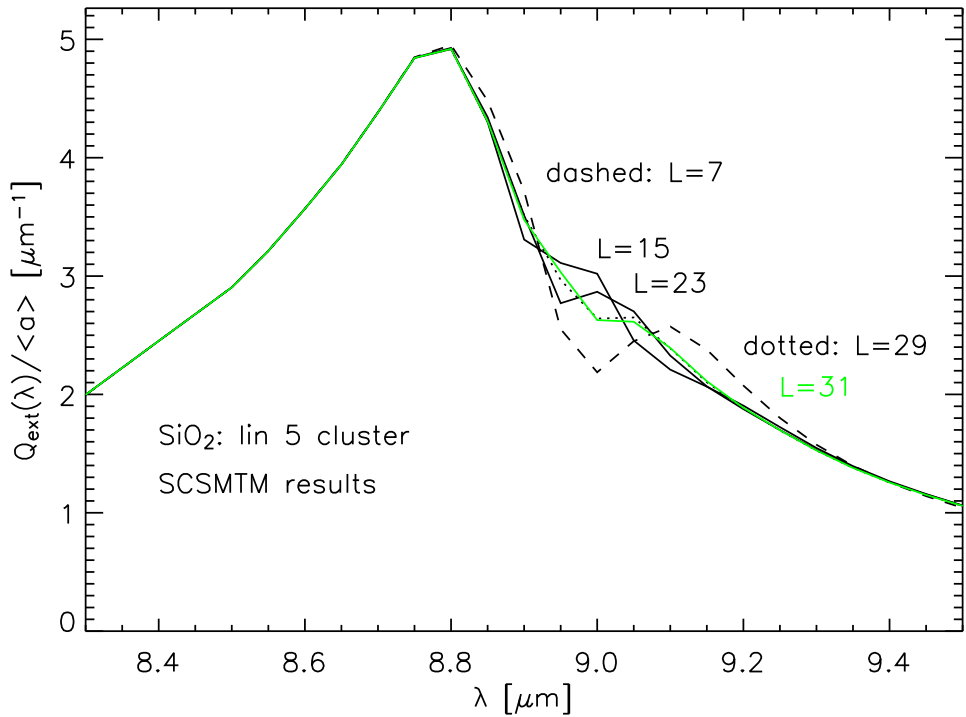}}
\caption{The calculated extinction of a SiO$_2$ linear five particles cluster {\it lin5} using the SCSMTM code \cite{mackowski96}. The polar order L shown spans from 7 to 31.}
\label{SiO2_lin5_SCSMTM}
\end{figure}

The clusters-of-spheres calculations have been performed using 
(1) the T-matrix code by D.W.\,Mackowski (SCSMTM) calculating the 
random-orienta\-tion scattering matrix for an ensemble of spheres
\cite{mackowski96} and (2) the program developed by M.\,Quinten \cite{quinten93}
(MQAGGR, commercially available) based on the theoretical 
approach by \cite{gerardy82}. 
Both programs aim at solving the scattering 
problem in an exact way by treating the superposition of incident and 
all scattered fields, developed into a series of vector spherical harmonics. 
Available computer power, however, forces to truncate the series at 
a certain maximum multipolar order L = npol$_{max}$, which in both programs 
can be specified explicitly. Furthermore, both programs perform an 
orientation average of the cluster, the resolution of which was 
set to 15 degrees in MQAGGR and 10 degrees in SCSMTM for theta 
(the scattering angle). The variation in the azi\-muthal angle is not 
specified in SCSMTM. In MQAGGR it is varied between 0 and 360 degrees, 
again with a resolution of 15 degrees. 

The discrete dipole approximation (DDA) method is one of several
discretization methods (e.g.\ \cite{draine88,hage+greenberg}) 
for solving scattering problems in the presence of a target with 
arbitrary geometry.  In this work we use the DDSCAT code version 6.1 
by B.T.\,Draine and P.J.\,Flatau \cite{draine+flatau04}, which is very popular among astrophysicists.
In DDSCAT the considered grain/cluster is replaced by a cubic array
of point dipoles of certain polarizabilities \cite{goodman}.
The cubic array has numerical advantages because it allows a significant
speedup of the conjugate gradient method, applied to solve the matrix
equations describing the dipole interactions, by using Fast Fourier Transforms.
By specifying an appropriate
grid resolution, calculations of the scattering and absorption of
light by inhomogeneous media such as particle aggregates can be carried out
to in principle whatever accuracy is required. The size of the dipole grid is in reality limited by the available computing power.

M.\,Min et al. \cite{min+etal05} have presented an easy to use method to 
compute the absorption and scattering properties of small particles with 
arbitrary shape, structure, orientation and composition based on a solution of 
the DDA equations in the Rayleigh domain. For a given geometrical shape of 
the particles, the solution has to be computed only once to obtain the 
absorption and scattering properties for arbitrary values of the refractive 
index. This method thus provides a nice tool to study the dependence of the convergence
behavior of the DDA on the value of the refractive index of the particle.

\section{Convergence of the clusters-of-spheres method}{\label{sect_cos}}

For a pair of gold spheres, it has been shown already by \cite{smith+etal95} that the
clusters-of-spheres method, when applied to touching particles, hardly yields
convergent results when including multipolar terms up to an order of eight.
By comparing their results -- obtained according to the formalism by \cite{gerardy82}
-- with the exact quasistatic solution for a pair of touching
particles, Smith et al.\ found that especially for the longitudinal optical mode,
the multipolar expansion up to L = 8 does not lead to satisfactory results.
(Note that the values of n and k for gold near the 484\,nm plasmon resonance
peak studied by \cite{smith+etal95} amount to n$_{res}$ = 1.1 and k$_{res}$ = 1.8.
Beyond $\lambda_{res}$, k($\lambda$) increases with $\lambda$.)

In our application of the clusters-of-spheres method, we came to similar
conclusions, even though we included multipolar terms up to an order of
L = 31. This is illustrated in Figs.\,\ref{lin5_SiC_SCSMTM} -- \ref{SiO2_lin5_SCSMTM}, 
which show the extinction band profiles for the {\it lin5} cluster of SiC, 
FeO and SiO$_2$ spheres in the 10--12.5, 15--32 and 8.3--9.5\,$\mu$m wavelength 
ranges, respectively.  The spectra are presented in terms of the extinction 
efficiency Q$_{ext}$=C$_{ext}$/($\pi$$<$a$>$$^2$), with C$_{ext}$ being the extinction 
cross section and $<$a$>$ the radius of a sphere with the same volume as the cluster. Q$_{ext}$ is 
normalized by $<$a$>$ to become independent of $<$a$>$ within the Rayleigh limit. In the figures we plot Q$_{ext}$/$<$a$>$ because this quantity is independent of the particle size when the particles are in the Rayleigh limit.

For all three materials, a phonon band -- split up into at least 
two components -- is seen in the extinction spectrum. Also in all three cases, 
for the `blue' component of the phonon band, convergence is achieved for L $>$ 20, 
while for the `red' component(s), this is not the case,
but instead, even for L = 31, there is a continuous peak shift if L is increased.
Recall that our extinction spectra do not refer to specific
orientations of the particle chain with respect to the external electromagnetic
field, but are orientationally averaged. By studying specific orientations of the
chains, it becomes evident that the `blue' phonon band component corresponds to
an orientation of the electric field perpendicular to the long axis of the chain,
while the `red' component corresponds to the complementary case. It is this
latter case which causes the slow convergence. The convergence, however, also
depends on the values of optical constants. For amorphous SiO$_2$, with its
smaller maximum values of n and k (cf.\ Tab.\,\ref{t:nk-max}), the differences between the
results for the individual multipolar orders are significantly smaller than
for FeO with its larger peak values of n and k.

\begin{table}
\begin{center}
         \caption{The number of dipoles used with the DDSCAT \cite{draine+flatau04} calculation as well as the number of dipoles actually within the cluster.
         \label{dipoles}} 
\begin{tabular}{|lcc|lcc|}
\hline
Cluster & $\#$ dipoles & $\#$ dipoles & Cluster & $\#$ dipoles & $\#$ dipoles   \\ 
&  in the grid & in the cluster &  & in the grid & in the cluster  \\ \hline \hline
{\it lin5}  & 48x16x16 = 12288 & 2316 & {\it sc8} & 24x24x24 = 13824 & 7160 \\
{\it lin5}  & 60x60x60 = 216000  & 4475 & {\it sc8} & 25x25x25 = 15625 & 8217 \\
{\it lin5}  & 72x48x32 = 110592  & 7792 & {\it sc8} & 60x60x60 = 216000 & 111976 \\
{\it lin5}  & 80x80x80 =  512000 & 10515  & {\it frac7} & 24x24x24 = 13824 & 1757 \\
{\it lin5}  & 90x60x40 = 216000 & 14845 & {\it frac7} & 25x25x25 = 15625 & 2105 \\
 {\it lin5}  &  96x48x24 = 110592 & 18620 & {\it frac7} & 60x60x60 = 216000 & 28973 \\ \hline 
\end{tabular} 
\end{center}
\end{table}

\begin{figure}
\resizebox{\hsize}{!}{\includegraphics{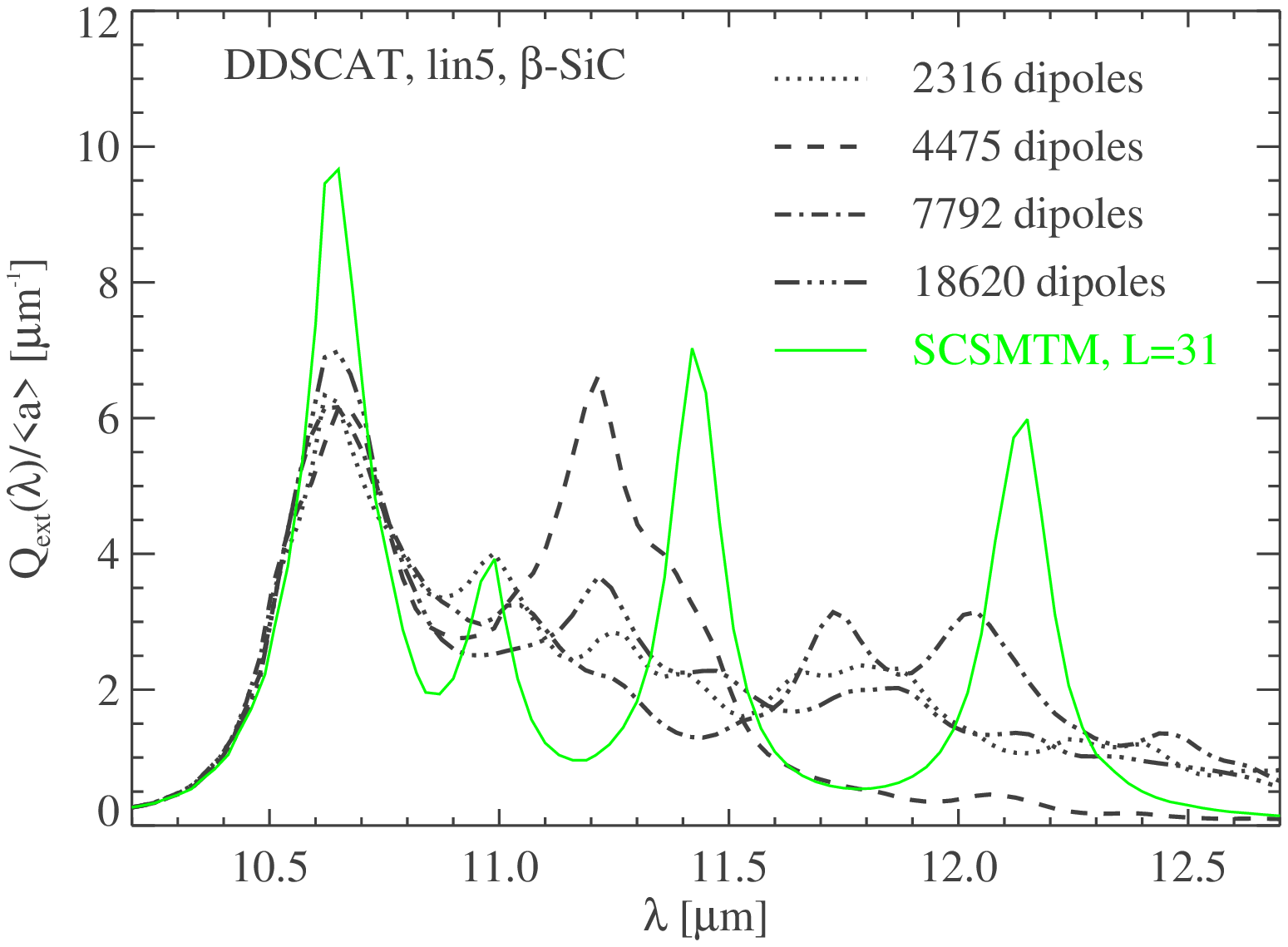}}
\caption{The calculated extinction of a SiC linear five particle cluster {\it lin5} using the DDSCAT code by \cite{draine+flatau04}. 
The number of the dipoles within the cluster is indicated by the annotations, see also Tab.\,\ref{dipoles}. For comparison the result as calculated with the SCSMTM code \cite{mackowski96} for L = 31 is shown.
}
\label{lin5_SiC_DDSCAT}
\end{figure}

\begin{figure}
\resizebox{\hsize}{!}{\includegraphics{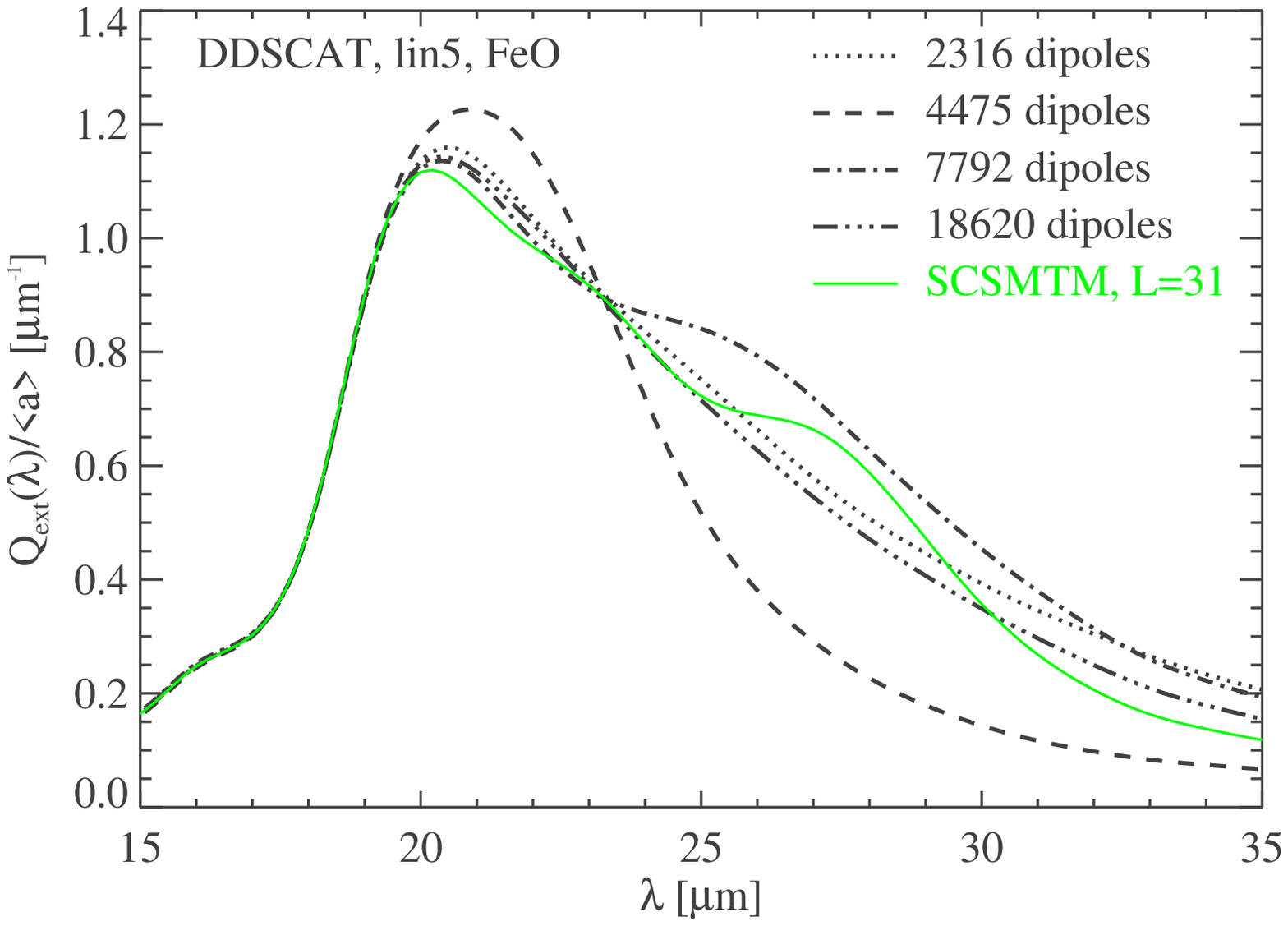}}
\caption{
The calculated extinction of a FeO linear five particle cluster {\it lin5}
using the DDSCAT code \cite{draine+flatau04}. The number of the dipoles within the cluster is indicated by the annotations, see also Tab.\,\ref{dipoles}. For comparison the result as calculated with the SCSMTM code \cite{mackowski96} for L = 31 is shown.
}
\label{FeO_Lin5_DDSCAT}
\end{figure}

\begin{figure}
\resizebox{\hsize}{!}{\includegraphics{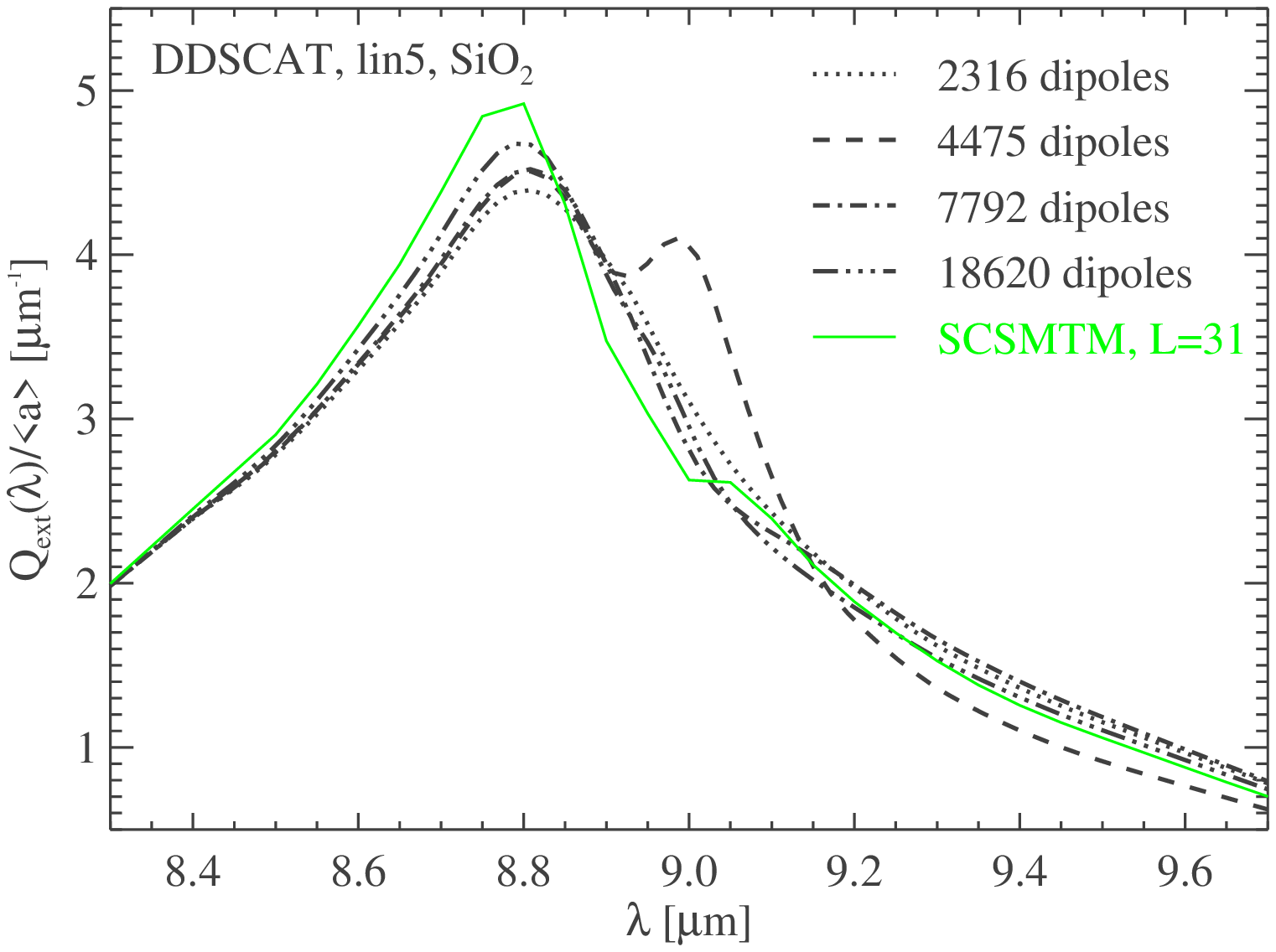}}
\caption{
The calculated extinction of a SiO$_2$ linear five particle cluster {\it lin5}
using the DDSCAT code \cite{draine+flatau04}. The number of the dipoles within the cluster is indicated by the annotations, see also Tab.\,\ref{dipoles}. For comparison the result as calculated with the SCSMTM code \cite{mackowski96} for L = 31 is shown.
}
\label{SiO2_Lin5_DDSCAT}
\end{figure}

\begin{figure}
\resizebox{\hsize}{!}{\includegraphics{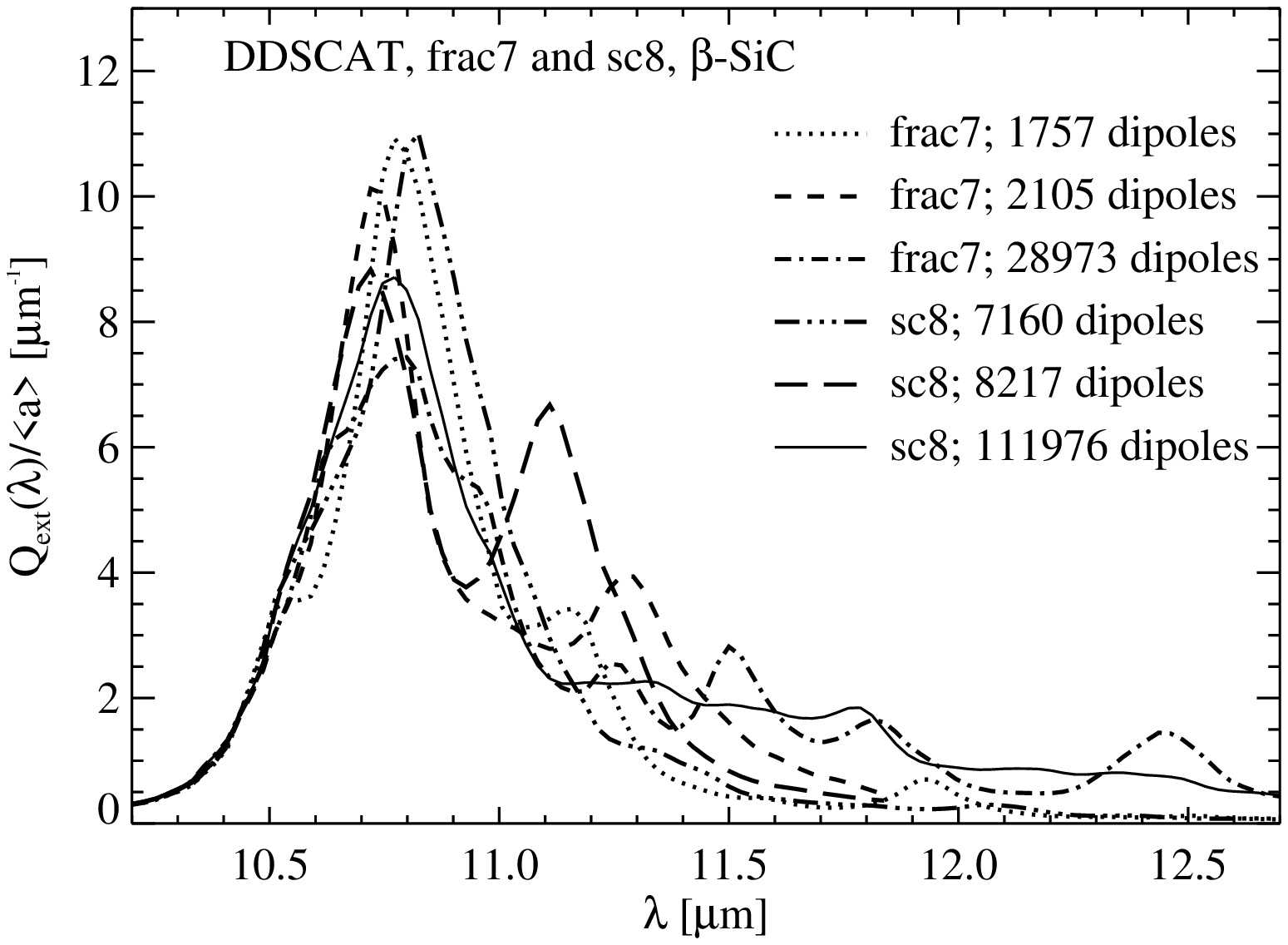}}
\caption{The calculated extinction of the SiC clusters {\it frac7} and {\it sc8} using the DDSCAT code \cite{draine+flatau04}. The number of dipoles within the cluster is indicated by the annotation, see Tab.\,\ref{dipoles} for details. 
}
\label{frac7sc8_SiC_DDSCAT}
\end{figure}


The 20\,$\mu$m phonon band of amorphous SiO$_2$ (not shown in our figures) 
actually converges even faster than the the 9\,$\mu$m band. Here, 
convergence seems to be achieved for L$<$31. This is probably due 
to the fact that Re(m) does not reach as low values as for the 
9\,$\mu$m band. This dependence of the convergence on Re(m) is also 
observed with the DDA approach and will further be discussed in the
next section. 

\section{Convergence of the DDA methods}{\label{sect_DDA}}

As well as the cluster-of-spheres methods, the DDA method also encounters difficulties 
with reaching convergence for the clusters we have chosen to study, 
see Figs.\,\ref{lin5_SiC_DDSCAT} -- \ref{frac7sc8_SiC_DDSCAT}.
For the {\it lin5} cluster composed of SiC it is clear from 
Fig.\,\ref{lin5_SiC_DDSCAT} that we are most 
likely not even close to obtaining a converged result, although we have used
rather large numbers of dipoles (see Tab.\,\ref{dipoles}) 
for the calculations. This is especially true at wavelengths longward of the 
primary resonance at 10.6~$\mu$m. Some of the spectra obtained show 
secondary resonances similar to the SCSMTM results whereas others are 
relatively flat. 
Similarly, for the FeO {\it lin5} cluster the results seem to be converged 
towards the short-wavelength part of the spectrum shown in Fig.\,\ref{FeO_Lin5_DDSCAT}, 
but not towards the longer wavelengths, especially for one particular spectrum 
(4475 dipoles) which strongly differs from all others including the SCSMTM result. 
For SiO$_2$ (Fig.\,\ref{SiO2_Lin5_DDSCAT}) the DDSCAT calculation is closer to 
converging due to the lower values of the optical constants for this material 
as compared to FeO and SiC, see Tab.\,\ref{t:nk-max}. However, again the 
4475 dipole spectrum shows a secondary resonance which is neither present in 
the other three nor in the SCSMTM result. 
Similar results concerning the convergence were found for the two other cluster 
shapes studied {\it frac7} and {\it sc8}, Fig.\,\ref{frac7sc8_SiC_DDSCAT}. 

An influence of the refractive index values on the precision of the DDA 
is expected. For materials with large refractive index ($|m| > 2$) 
\cite{draine+goodman} show that especially the absorption is overestimated
by DDA.  
According to \cite{draine+goodman}, the lattice dispersion relation 
prescription for the polarisability, $\alpha_{i}$, gives fair accuracy for scattering but poorer
results for absorption. When $|m - 1|$ is large, the continuum material
is effective at screening the external field:
in the limit $|m - 1| \rightarrow \infty$
the internal field generated by the polarization would exactly cancel the
incident field, so that the continuum material in the interior of the target
would be subjected to zero field. In the case of a discrete dipole array,
the dipoles in the interior will also be effectively shielded, while
the dipoles located on the target surface are not fully shielded and,
as a result, absorb energy from the external field at an excessive rate.
This error can be
reduced to any desired level by increasing the number $N$ of dipoles,
thereby minimizing the fraction $~ N^{-1/3}$ of the dipoles which are at
surface sites, but very large values of $N$ are required when $|m - 1|$ is
large \cite{draine+goodman}.

\begin{figure}
\centering\includegraphics[width=9cm,angle=0]{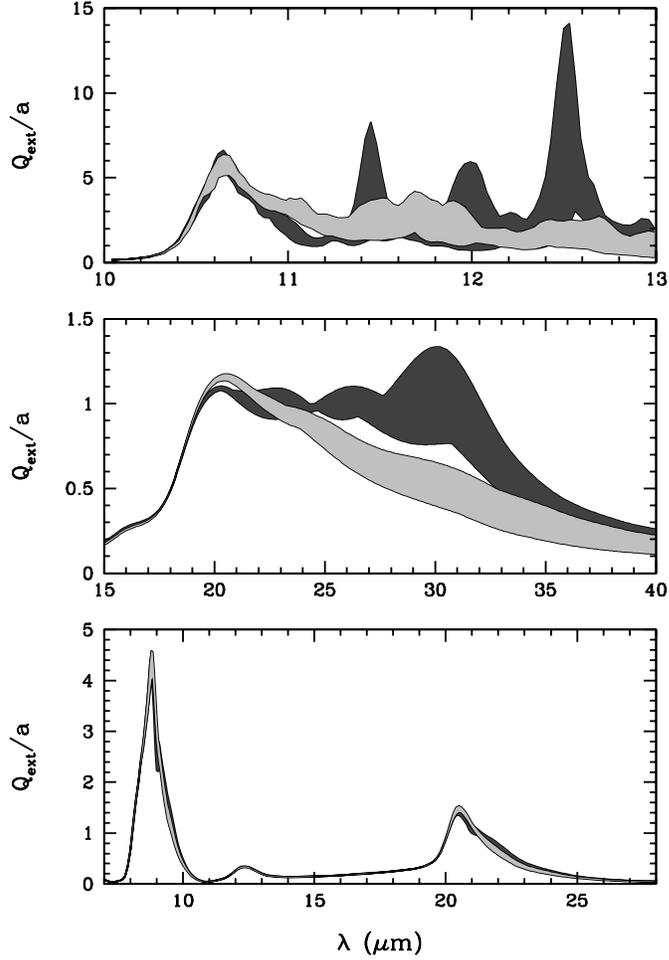}
\caption{A demonstration of how the results of the DDA calculations for the {\it lin5} cluster show a dependency on the choice of dipole grid used. The top panel shows the SiC cluster, the middle panel the FeO cluster and the bottom panel the SiO$_2$ cluster. The results are obtained with the code by \cite{min+etal05}, an identical dependence is found by the DDSCAT code of \cite{draine+flatau04}. 
The dark gray area show the range of seven spectra obtained when using a dipole grid which
is a multiple of five. The light gray shaded area is the range of 24 spectra obtained when using 
dipole grid which isn't a multiple of five. A similar result is found for the {\it frac7} and
{\it sc8} clusters but here for grids which are multiple of three and two, respectively.
}
\label{multiple}
\end{figure}

At a closer view to the results of the DDA calculations, we found that 
the choice of the dipole grid is significant for the obtained result. 
For the linear five particle cluster {\it lin5}, 
a grid with a dipole number which is a multiple of five provides a 
significantly different result than dipole grids which are not
a multiple of five. The former are the ones which generally produce 
stronger secondary resonances as can be seen from a comparison of the 
resulting spectra (shown in Fig.\,\ref{lin5_SiC_DDSCAT}) with Tab.\,\ref{dipoles} (see also Fig.\,\ref{multiple}, which will be discussed later). For the semi-fractal 
seven particle cluster {\it frac7} the difference was between dipole grids 
which were a multiple of  three and for the cubic eight cluster {\it sc8} 
it was for dipole grids which were a multiple of two, equivalent to the 
length scale of the longest chain within the cluster.

The results shown in 
Figs.\,\ref{lin5_SiC_DDSCAT} -- \ref{frac7sc8_SiC_DDSCAT} were calculated using 
DDSCAT \cite{draine+flatau04} but identical results were found using the code
by \cite{min+etal05}. According to this solution the absorption cross section 
averaged over all orientations of an arbitrarily shaped particle obtained using DDA 
can be written as
\begin{equation}
\label{eq:Cabs}
C_\mathrm{abs}=\sum_{j=1}^{3N} \frac{w_j}{N}~\left[kV~\mathrm{Im}\left(\frac{m^2-1}{1+L_j~(m^2-1)}\right)\right],
\end{equation}
where $k=2\pi/\lambda$, $V$ is the material volume of the particle, $N$ is the 
number of dipoles. The $0>L_j>1$, the so-called form-factors (not to be confused with the symbol for the multipolar order), and $w_j$, the 
weights, are obtained from the DDA equations. Once the $L_j$ and the $w_j$ are found, 
it is trivial to compute the absorption cross section for any wavelength or 
complex refractive index, $m$. 

\begin{figure}
\resizebox{\hsize}{!}{\includegraphics{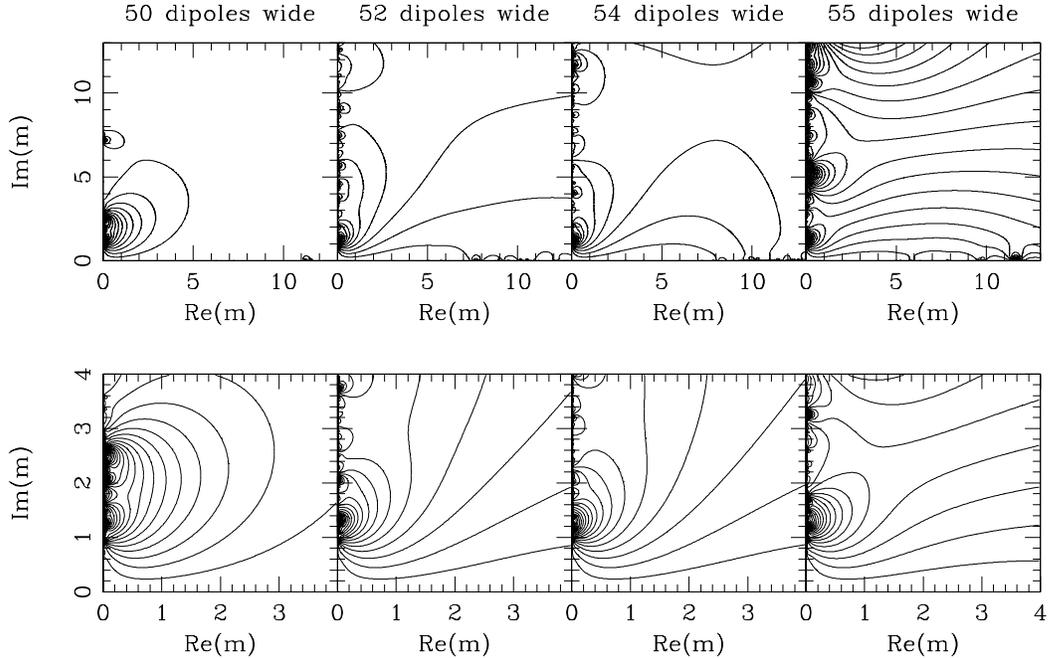}}
\caption{Contour plots of $C_\mathrm{abs}/(kV)$ as a function of the real and imaginary part of the refractive 
index for the {\it lin5} cluster, averaged over all particle orientations. The contours are obtained using DDA with 
a dipole grid which has 50 (left panels), 52 and 54 (middle panels) and 55 (right panels) dipoles along the long 
axis of the system. The lower panels are simply blowups of the lower left corners of the upper panels. The contours are plotted with increments of $0.5$.}
\label{fig:contours}
\end{figure}

Fig.\,\ref{multiple} shows a selection of results for the {\it lin5} cluster, 
divided in into grids dividable and non-dividable by 5. 
It is obvious, especially for SiC and FeO, that the dividable grids produce 
secondary resonances, which lead to long-wavelength extinction enhancement 
but with a large spread. Consequently, they are the ones that contribute 
stronger to the convergence problems than the non-dividable grids. 

The method by \cite{min+etal05} allows to examine these effects in even more 
detail by investigating the origin of the resonances. According to 
Eq.\,\ref{eq:Cabs}, each form-factor $L_j$ causes a resonance in the 
complex $m$ plane at
\begin{equation}
m=\sqrt{1-\frac{1}{L_j}}.
\end{equation}
In the limiting 
case $N\rightarrow\infty$ the $L_j$ and $w_j$ form a continuous distribution, 
smoothing out all these resonances to a continuum function of $m$. However, using a 
limited number of dipoles will result in a limited number of resonances. 
Especially for small values of $\mathrm{Re}(m)$, this resonance behavior might 
show up. The positions of these resonances are sensitive to the exact configuration 
of the dipoles in the particle. Thus it is very hard to obtain convergence using 
DDA for low values of $\mathrm{Re}(m)$.

The previous is illustrated in Fig.\,\ref{fig:contours} where we plot in the 
contours the dimensionless quantity $C_\mathrm{abs}/(kV)$ for the {\it lin5} 
aggregate as a function of the real and imaginary part of the refractive index. 
We do not give any numerical values for the contour lines because we would 
like to focus on the general shape of the contours. In the upper panels it 
can be seen that there is a large difference between the contours obtained 
using different numbers of dipoles. It is clear in all plots that a number 
of resonances are present along the axis with $\mathrm{Re}(m)=0$. More 
generally speaking, the contours look very different over the whole complex $m$ space. 
If we zoom in on the region with moderate values of $m$ (the lower panels), 
we see that the differences between the 52 and the 54 wide dipole grids are 
almost gone in this region of the complex $m$ plane. All the differences 
between the contours shown in the middle two panels are located at very 
small values of $\mathrm{Re}(m)$. However, the grid of 50 and 55 dipoles 
width still displays a different shape of the contours. More generally, 
for this cluster shape there is a different behaviour when the number of 
dipoles along the long axis of the system is a multiple of five. 
This illustrates the bad convergence behavior of this 
cluster shape when using DDA.

\section{Conclusions}\label{conclusion_sect}

We have studied the performance of two cluster-of-spheres (\cite{mackowski96,quinten93}) and two discrete dipole approximation (DDA; \cite{draine+flatau04,min+etal05}) methods for calculating the extinction of aggregates. 
The methods were chosen because they are popular and often used in studies of 
astrophysical dust extinction and scattering problems.

We present results of the calculated extinction within infrared absorption bands 
of SiC, FeO and SiO$_2$ clusters composed of 5 to 8 spherical particles of radii 10 nm. 
The clusters had three different shapes: linear chain, semi-fractal and simple cubic.
The materials display a range of material properties and therefore have different
strengths of the surface resonances. When the real part of the 
refractive index is much smaller than unity, none of the four methods are able
to converge. For the two DDA methods there is a strong dependence of the calculated
band profiles on the exact dipole distribution within the particles. For the 
linear five particle cluster {\it lin5} the result depend on whether
the grid is a multiple of five or not, for the semi-fractal seven particle
cluster {\it frac7} and the simple cubic eight particle cluster {\it sc8} the
dependence is on whether the grid is a multiple of three and two, respectively.

Currently it does not seem possible to 
calculate the absorption efficiencies of clusters of spheres for materials with optical constants $n<<1$ in a
reliable way. We assume that the critical point is the contact geometry of the particles, which
explains the importance of the resolution of the methods used. This implies that 
for corresponding experiments it is of great importance to study the 
contact points
between the particles. Without a good investigation of the contact points a 
comparison between experimental data and model calculations are very likely to be fruitless. One could hope that things might be easier when particles are less perfect than the spheres we have considered, since that will give rise to fewer
resonance effects.

\subsection*{Acknowledgment}
      We would like to thank B.T.\,Draine and P.J.\,Flatau for making their
      DDA code and D.W.\,Mackowski for making his T-matix code available 
      as shareware.
      HM acknowledges support by DFS grant Mu1164/5, 
      MM would like to thank J.W.\ Hovenier for valuable discussions and 
      TP acknowledges R.\ Ottensamer's significant contributions to the SCSMTM calculations. 
      Dark Cosmology Center is funded by the Danish National Research Foundation.

\end{document}